\documentclass[12pt,a4paper]{article}

\usepackage{a4}

\usepackage{amsfonts}
\usepackage{amssymb}
\usepackage{epsfig}
\usepackage{psfig}
\usepackage{psfrag}
\usepackage{dsfont}

\usepackage[sc,small]{caption2}
\setcaptionwidth{14cm}

\newcommand{\beqn}{\begin{eqnarray}}
\newcommand{\eeqn}{\end{eqnarray}}
\newcommand{\be}{\begin{equation}}
\newcommand{\ee}{\end{equation}}
\newcommand{\non}{\nonumber \\}

\newcommand{\tn}{{\tilde \nu}}

\newcommand{\ca}{{\cal A}}
\newcommand{\cn}{{\cal N}}

\newcommand{\cf}{{\cal F}}
\newcommand{\cv}{{\cal V}}

\newcommand{\ck}{{\cal K}}

\newcommand{\zba}[2]{[\!\!\begin{array}{c}{\scriptstyle#1}%
                        \\[-1.6mm]{\scriptstyle #2}\end{array}\!\!]}
\newcommand{\tht}{\vartheta}
\newcommand{\thb}[2]{\vartheta[{^{\, #1}_{\, #2}}]}

\newcommand{\thba}[2]{\vartheta[\!\!\begin{array}{c}{\scriptstyle#1}%
                        \\[-1.6mm]{\scriptstyle #2}\end{array}\!\!]}

\begin{document}

\title{}
\author{}
\date{}
\thispagestyle{empty}

\begin{flushright}
\vspace{-3cm}
{\small MIT-CTP-3542 \\
        NSF-KITP-04-110 \\
        hep-th/0409282}
\end{flushright}
\vspace{1cm}

\begin{center}
{\Large\bf 
On the Moduli Dependence of \\[0cm]
Nonperturbative Superpotentials \\[.3cm] in Brane Inflation 
}
\end{center}

\vspace{.5cm}

\begin{center}

{\bf Marcus Berg}$^{\dag}$
{\bf,\hspace{.2cm} Michael Haack}$^{\dag}$
{\bf\hspace{.1cm} and\hspace{.2cm} Boris K\"ors}$^{*}$
\vspace{0cm}
\end{center}

\hbox{
\parbox{7cm}{
\begin{center}
{\it
$^{\dag}$Kavli Institute for Theoretical Physics \\
University of California \\
Santa Barbara, \\
California 93106-4030, USA\\
}
\end{center}
}
\hspace{-.5cm}
\parbox{8cm}{
\begin{center}
{\it
$^*$Center for Theoretical Physics \\
Laboratory for Nuclear Science \\
and Department of Physics \\
Massachusetts Institute of Technology \\
Cambridge, Massachusetts 02139, USA \\
}
\end{center}
}
}

\vspace{1cm}

\begin{center}
{\bf Abstract} 
\end{center} 

We discuss string corrections to the effective potential in various models of
brane inflation. These corrections contribute to the mass of the inflaton
candidate and may improve its slow-roll properties. In particular, 
in orientifold string compactifications
with dynamical D3- and D7-branes, the corrections induce inflaton dependence
in the part of the superpotential that arises from gaugino condensation or
other nonperturbative effects. The additional terms are in part
required by supersymmetry. We explicitly discuss
D3/D7-inflation, where flat directions of the potential can be lifted, and
the KKLMMT model of warped brane inflation, in which the corrections 
open up the possibility of flattening 
the potential and canceling unwanted 
contributions to the inflaton mass.

\clearpage
\setcounter{page}{1}


\section{The inflaton mass problem and the rho problem} 

While cosmological inflation is a very attractive mechanism to solve
many of the inherent problems of standard big bang cosmology, there
are still many open questions about how it can be realized within
string theory. Let us single out two obstacles that played an
important role in the development of the models  considered
below. 
Firstly,
the no-go theorems of \cite{deWit:1986xg} show that it is
not possible to achieve a compactification of ten- or
eleven-dimensional supergravity with a positive cosmological constant
in four dimensions, at least at leading order in an expansion in
derivatives and in the string coupling. This conclusion is not changed even by
including orientifold planes, non-dynamical objects of formally
negative energy density, in the background \cite{Giddings:2001yu}.
The way out of this problem, which in its present concrete form was
proposed by Kachru, Kallosh, Linde and Trivedi \cite{Kachru:2003aw},
is to add nonperturbative effects of some sort.  The resulting
effective superpotential is our main concern in this note. 
The second obstacle, the
``inflaton mass problem'', sometimes also called the ``eta problem'', 
can be seen in 
terms of purely four-dimensional $\cn=1$ effective supergravity, 
 which says that the standard form of
the F-term potential,
\beqn 
\cv^F = e^\ck ( \ck^{I\bar J} D_I W D_{\bar J} \bar W - 3|W|^2 ) = e^\ck \tilde \cv^F \ , 
\eeqn 
given an expansion of the 
K\"ahler potential $\ck = \phi\bar\phi +\, \cdots$ 
in the fluctuations of the inflaton  
around $\langle \phi\rangle =0$, produces 
\beqn \label{etaprob}
\frac{\cv^F_{\phi\bar\phi}}{\cv^F} = 1 + \frac{\tilde\cv^F_{\phi\bar\phi}}{\tilde\cv^F}
~~ \Rightarrow~~ \eta \sim {\cal O} (1)\ , 
\eeqn 
i.e.\ a contribution of order one to the slow roll parameter 
$\eta$ \cite{Copeland:1994vg}.\footnote{Since the 
other slow-roll parameter $\epsilon$ is usually much smaller
 than $\eta$ in the models we consider, it will be ignored in the following.}
Obviously, this can be avoided by invoking a cancellation 
between the two terms in (\ref{etaprob}), 
moderately fine-tuning (at the level of $1\%$) 
to achieve slow-roll at a comfortable $\eta \sim 0.01$.
Still, 
this means slow-roll inflation can be realized but is not a generic property. 
One might then ask, 
given the fact that a supergravity potential generated 
by spontaneous supersymmetry breaking has 
an inflaton mass problem, 
could one use explicit supersymmetry 
breaking within string theory, 
e.g.\ 
anti-branes in otherwise supersymmetric 
backgrounds \cite{Dvali:1998pa}? 
Then the brane moduli, such as their positions, are
interpreted as the scalar inflaton fields \cite{Dvali}.  
Unfortunately, the simplest version of the brane/anti-brane scenario 
(in an unwarped background) produces a
classical Coulomb brane/anti-brane
attraction that is too strong to allow for slow-rolling, 
again $\eta \sim {\cal O} (1)$, and the inflation mass
problem from supergravity recurs.

There are, in fact, various ways to get around the
inflaton mass problem within brane inflation. We will focus on
two approaches here: 
One is the setting of D3/D7-brane inflation 
\cite{Dasgupta:2002ew,Hsu:2003cy,Hsu:2004hi,Dasgupta:2004dw},
which starts from an orientifold compactification 
with D3- and D7-branes and $\cn=2$ supersymmetry, 
then breaks supersymmetry completely by a small brane 
misalignment. This is 
described by an effective Fayet-Iliopoulos term 
(or better, a triplet of such terms). Thus the scenario is 
similar to D-term inflation \cite{Binetruy:1996xj} 
and therefore escapes the inflaton mass problem. 

The other approach is the model of 
Kachru, Kallosh, Linde, Maldacena, McAllister 
and Trivedi \cite{Kachru:2003sx}, who 
suggested that the brane/anti-brane
Coulomb attraction can be weakened through 
gravitational redshift, if the 
background metric includes a warp factor which differs at 
the position of the brane and the 
anti-brane by a sufficiently large amount. 
The KKLMMT model contains the following ingredients:
\begin{enumerate}
\item
The model is an orientifold compactification of type IIB on a 
(warped) Calabi-Yau 
with background RR and NSNS 3-form fluxes \cite{Giddings:2001yu}. 
\item
The warping is such that there are regions 
with very strong warping, usually modeled by the 
Klebanov-Strassler geometry \cite{Klebanov:2000hb}. 
\item
As a prototype of a nonperturbative effect,
one considers strongly coupled gauge dynamics on D7-branes 
wrapped on 4-cycles of the Calabi-Yau. 
\item
The inflaton 
scalar field arises as the position field of dynamical 
D3-branes. 
\item
Supersymmetry 
is broken either through anti-D3-branes 
\cite{Kachru:2003sx} or through non-BPS world-volume 
gauge field backgrounds on the D7-branes 
\cite{Burgess:2003ic}. 
In the case of
anti-D3-branes, it is assumed that the 
gravitational redshift is big enough to allow neglecting 
the Coulomb interaction in the potential.
\end{enumerate}

The models have many similarities, but also differ in important ways.  
In particular, points 2 and 3 of our list do not play any major role
in the model of \cite{Dasgupta:2002ew}, 
as the relevant brane interaction in that model
occurs within a single D3/D7-system (cf.\ fig.\ 4 of 
\cite{Dasgupta:2002ew}).\footnote{We thank R.\ Kallosh for pointing this 
out to us.} This requires a sufficient separation of the single 
D3/D7-system from the rest of the D7-branes. More generally, strong gauge
dynamics on these D7-branes will also play a role in $\cn=2$ models
of D3/D7-inflation and the results below are in general 
equally relevant for that case.

The potential of the KKLMMT model is
assembled from the following contributions. 
The action for the bulk Calabi-Yau was 
derived through generalized dimensional 
reduction and related to gauged supergravity 
(see e.g.\ \cite{Kachru:2002he,DeWolfe:2002nn}). 
The RR and NSNS 3-form fluxes 
add a superpotential of the form \cite{Gukov:1999ya}
\beqn 
W_{\rm 3-flux} = \int_{\rm CY} \Omega_3 \wedge ( F_3 - \tau H_3) \ , 
\eeqn 
which depends on the complex dilaton 
$\tau$ and the complex structure moduli $u^I$. In general, the 
supersymmetry conditions $D_\tau W= D_{u^I} W=0$ can be strong 
enough to fix all these fields. 
It is then assumed that their masses upon stabilization become 
large enough that $\tau$ and $u^I$ can 
be completely ignored during inflation 
(also, the string coupling should be fixed in the perturbative regime). 
Furthermore, as was discussed in \cite{Gorlich:2004qm}, 
one can argue that the open string moduli 
of D7-branes are fixed at the same mass scale. Then,  
the only relevant degrees of freedom below this scale
are the K\"ahler moduli 
and the D3-brane scalars, and 
in the simple case of only one K\"ahler modulus $\rho$, one may
 restrict to a model of 
only two fields: the K\"ahler modulus $\rho$ and the inflaton $\phi$. 
Gaugino condensation on the D7-branes 
induces a nonperturbative superpotential
involving the gauge kinetic function $f_7$ of the D7-brane 
world volume gauge theory, 
\beqn \label{wtotal}
W ~=~ W_{\rm 3-flux} + W_{\rm nonpert} ~\sim~ W_0 + C e^{-a f_7} 
\eeqn 
for some constants $a,C$ and $W_0$, where $W_0$ 
is the value of the flux-induced 
superpotential at the minimum of the potential.\footnote{The notation
in (\ref{wtotal}) is not meant to imply that
$W_{\rm nonpert}$ cannot also depend on the 3-form fluxes. The 
gauge kinetic function $f_7$ may receive corrections 
in their presence.}
A priori, $f_7$ is a holomorphic function of all the chiral fields, 
$f_7 = f_7(\rho,\phi)$. Now one can find supersymmetric vacua 
with negative cosmological 
constant by imposing $D_\rho W=0$, with stabilized (and moderately 
large) volume. 
The K\"ahler potential for this $\rho\, $-$\,\phi$ model has 
been conjectured to be of the general 
form \cite{DeWolfe:2002nn} 
\beqn 
{\cal K} = -3 \ln [-i(\rho-\bar\rho) + k(\phi,\bar\phi)]
\eeqn
to leading order, 
where $k(\phi,\bar\phi)$ is the K\"ahler potential of the Calabi-Yau 
manifold. For small $\phi$ it can be approximated as 
$k(\phi,\bar\phi) = \phi\bar\phi + \,\cdots$, and we ignore the higher terms 
in what follows. 
It is important that the argument of the logarithm is still 
equal to the Calabi-Yau volume in the Einstein frame 
even when $\phi\not=0$, i.e.\ 
\beqn \label{rho}
-i(\rho-\bar\rho) + \phi\bar \phi  ~=~ e^{-\Phi} v^{2/3}\ , 
\eeqn 
where $v$ is the volume in the string frame. 
Eq.\ (\ref{rho}) implies a mixing of geometrical closed string 
moduli ($v$) and open string moduli ($\phi$) 
in the K\"ahler coordinate $\rho$. 
This mixing gives rise to two problems:
the ``rho problem'' 
and the inflaton mass problem, which we shall return to in a moment. 
Finally,
the last contribution to the potential that was added in \cite{Kachru:2003sx}
is the one due to anti-D3-branes
(that are non-dynamical in the presence of imaginary self-dual 3-form flux 
\cite{Kachru:2002gs})
which is written as a ``warped FI-term'':
\beqn \label{antiD}
 e^{-\Phi} \sqrt{ -{\rm P}[g_4]} 
\quad \stackrel{\mbox{\tiny Einstein frame}}{\longrightarrow}\quad
 \cv^D = \frac{ D}{[-i(\rho-\bar\rho)+\phi\bar\phi]^2}\; ,
\eeqn
where $D$ is a constant given by the 3-form fluxes and 
the tension of the anti-branes \cite{Kachru:2003sx}. 
This can be derived by reducing the Born-Infeld action in 
the warped background, the constant $D$ being 
the warp factor at their position, at the tip of the warped throat. 
If we ignore the dependence of $f_7$ on $\phi$, i.e.\
if we set $f_7=f_7(\rho)$,
we have an inflaton mass problem 
in the model. Indeed, the full F-term potential reads 
(with $W_\rho = \partial_\rho W$, etc.)
\beqn
\cv^F &=& \frac{1}{3[-i(\rho-\bar\rho)+\phi\bar\phi]^2} \Big( - i(\rho-\bar\rho) |W_\rho|^2 
\non
&& 
\hspace{3.5cm} 
- 
 3i( W_\rho \bar W - {\rm c.c.}) - |W_\phi|^2 - 
i(\phi W_\phi \bar W_{\bar \rho} - {\rm c.c.}) \Big)\ . 
\eeqn
Then, for $f_{7}=f_{7}(\rho)$ or $W = W(\rho)$, the potential $\cv = \cv^F + \cv^D$ 
depends on $\phi$ only 
through the prefactor $[-i(\rho-\bar\rho)+\phi\bar\phi]^{-2}$ and it 
follows\footnote{Note that $\eta \sim {\cal O}(1)$
becomes obvious only after 
normalizing the inflaton field correctly, i.e.\ after introducing
the canonically normalized inflaton 
$\varphi \sim \phi/\sqrt{-i(\rho-\bar \rho)}$ \cite{Kachru:2003sx}.}  
\beqn \label{etaprobl}
\cv_{\phi\bar\phi} \; \propto \; \cv \quad \Rightarrow \quad 
\eta \sim {\cal O}(1) \; . 
\eeqn 
There is nothing to tune,
so the inflaton mass problem looks
incurable at this level.
It was also argued that this mass term 
is related to the conformal coupling on the 3-brane world-volume and 
thus (\ref{etaprobl}) was to be expected \cite{Kachru:2003sx,Buchel:2003qj}. 
Fortunately,  there are contributions to $\eta$ that we 
have  neglected until now, and 
they may be useful for canceling  the above
terms. Indeed,  $f_7$ may depend on $\phi$ :
\beqn 
f_7 = f_{7}(\rho,\phi) \quad
\Rightarrow \quad 
W_{\rm nonpert} \sim e^{-af_{\rm 7}} =  w(\phi) e^{ia\rho}\ . 
\eeqn
Computing the function $w(\phi)$ in a controllable, sufficiently
simple model was 
the goal of \cite{Berg:2004ek}.\footnote{There have been alternative 
approaches, such as \cite{Iizuka:2004ct}, where 
a cancellation among the above mass contribution and the 
Coulomb interaction was designed. However, such a cancellation requires
rather small warping, because the Coulomb interaction 
between the D3-branes and anti-D3-branes is suppressed by the warp factor. 
Thus it is a setup slightly different from that considered here.} 

\begin{figure}[h]
\begin{center} 
\hspace{0cm}
{
\psfrag{phi}[bc][bc][.7][0]{$\phi$}
\psfrag{F}[bc][bc][.7][0]{${\cal F}^{\mu\nu}_{\rm 7}$}
\psfrag{D3}[bc][bc][.7][0]{D3}
\psfrag{D7}[bc][bc][.7][0]{D7}
 \epsfxsize=10cm
 \epsfysize=3cm
\epsfbox{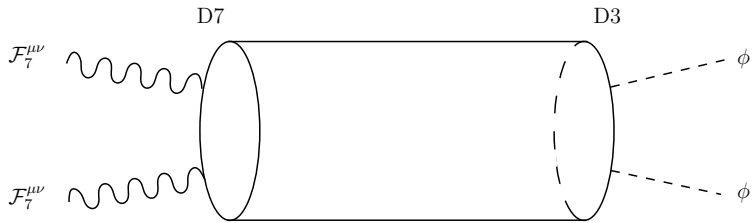}
\caption{The 3-7 annulus} \label{figur}
}
%
\end{center} 
\end{figure}

To understand
how the dependence of $f_7$ on $\phi$ arises, it is very useful to 
consider the following puzzle, the ``rho problem'',
and its resolution. 
The gauge coupling constant on the D7-branes is at leading 
order given by the dimensional reduction 
of the Born-Infeld action through 
\beqn \hspace{-.5cm}
e^{-\Phi} {\rm tr} \sqrt{ -{\rm det}({\rm P}[G]+{\cal F}_{\rm 7})} &\sim&
\underbrace{e^{-\Phi} v^{2/3}}_{=-i(\rho-\bar\rho) + \phi\bar\phi} 
 \!\!\!\sqrt{-g_4}\,\, {\rm tr}\, {\cal F}_{\rm 7}^2 \ . 
\eeqn
By supersymmetry, the gauge coupling must be the real part of a holomorphic
function, 
but it seems that it is not! 
Clearly,
$-i(\rho-\bar\rho)+\phi\bar\phi$ is not the real part of any function 
holomorphic in $\rho$ and $\phi$.
Thus, it looks 
as if we are facing a violation of supersymmetry by  violation 
of holomorphy \cite{Hsu:2003cy}. 
Two observations hint towards the solution: $i)$ For a stack of D3-branes, 
the field $\phi$ transforms in some 
representation of the D3-brane gauge group, and thus $\phi\bar\phi$ 
should really be read as 
$\phi^a\bar\phi^a={\rm tr}\, \phi\bar\phi$. Then, the term 
tr $\phi\bar\phi\, {\rm tr}\,  {\cal F}_{\rm 7}^2$ carries 
two traces, and has to come from a string diagram with at least 
two boundaries. $ii)$ At $\phi=0$, the 
volume modulus $\rho$ contains the factor $e^{-\Phi}$ whereas 
tr $\phi\bar\phi$ does not, and so the 
two terms arise at different orders
 of string perturbation theory; one from the disk, the other from 
open string one-loop level (annulus plus M\"obius strip). 
In all, the suspicion arises that
there could be a one-loop contribution 
from an annulus diagram with two boundaries, 
one on the D7- the other on the D3-branes, which may produce 
just the right dependence of the 
D7-brane gauge coupling to reinstate holomorphy and 
hence supersymmetry, cf.\ fig.\ \ref{figur}. 

This argument turns out to be correct \cite{Berg:2004ek}. 
In fact, the rho problem, that we only
invoked as a clue towards addressing the inflaton mass problem,
is more general than the application 
in the cosmological model at hand, and 
arises in any compactification that includes mobile D-branes.
In addition, the same 3-7 annulus diagram leads to the above 
mentioned dependence of the superpotential on the inflaton field
$\phi$ after gaugino condensation. 

We will not repeat any details of the computations of 
\cite{Berg:2004ek}, only 
discuss and interpret the results. 
We do add one ingredient 
relative to \cite{Berg:2004ek}: an alternative 
and independent calculation of the relevant terms using
standard string perturbation theory instead of the background field
formalism. 
This provides an independent confirmation 
of our earlier results and is an independent check of
 the validity of the background field
method in the problem at hand.   
The calculation is given in appendix \ref{appA}.


\section{On D3/D7-brane inflation} 

In this section we comment on the simplest setting in which the corrections 
to the gauge kinetic functions of D7-branes play a role, 
the model of D3/D7-brane inflation. 
As this model has more moduli than $\rho$ and $\phi$, it will not be sufficient to use the 
$\cn=1$ notation of the previous section, and we need more general 
expressions for 
the effective Lagrangian in this section only. In the next
section on warped brane inflation with $\cn=1$ 
supersymmetry, we shall switch back to $\rho$ and $\phi$, 
essentially by truncating the  $\cn=2$ model. 

In order to explicitly 
calculate the relevant one-loop diagrams 
shown in figure \ref{figur}, 
we use the simplest available orientifold model 
with D3- and D7-branes: the 
$\cn=2$ type IIB orientifold on  
$\mathbb{T}^2 \times {\rm K3} 
= \mathbb{T}^2 \times \mathbb{T}^4/\mathbb{Z}_2,$ 
where the $\mathbb{Z}_2$ is the reflection along the 
four circles \cite{Bianchi:1990yu,Angelantonj:2002ct}, 
and the orientifold projection is the T-dual of the world 
sheet parity $\Omega$, i.e.\ 
$\Omega'=
\Omega R^6(-1)^{F_L}$, where $R^6$ is the reflection 
along all six coordinates.\footnote{The results of \cite{Berg:2004ek} 
were mostly given 
in the T-dual picture with D9/D5-branes, but are unaffected by this 
T-duality, up to the standard transformation of the fields. In 
particular, all amplitudes computed there for D9/D5-branes are identical for 
D3/D7-branes 
after exchanging momentum and winding modes. Our alternative calculation 
 in the appendix is performed directly in
D3/D7-brane language, and  there we also check 
the T-duality mapping.}
This is precisely the model discussed in the context of 
$\cn=2$ D3/D7-inflation \cite{Dasgupta:2002ew}, before 
adding the deformation that breaks supersymmetry. 
The $\cn=2$ gauged supergravity Lagrangian of this compactification, 
including the 3-form RR and NSNS fluxes, was discussed in 
depth in \cite{Angelantonj:2003zx,D'Auria:2004qv}. 
The model has 32 D3-branes and 32 D7-branes with 
maximal gauge group $U(16)_9\times U(16)_5$, the D7-branes wrapping 
the K3. 
Here we are
 only interested in the vector multiplet sector of the full theory,
but including the branes and their open string degrees of freedom. In
total, there are three vector multiplets from the KK reduction of the
bulk fields and the vector multiplets that arise from the D3- and
D7-branes, which in general carry a representation of their
non-abelian gauge groups. In the abelian limit, where this group is
broken through an adjoint Higgs mechanism, these can be replaced by
$16+16$ abelian vector multiplets. The scalars in the bulk multiplets
will be denoted $S, S\, '$ and $U$, and the open string scalars $A_i$ for
D3-branes and $A_a$ for D7-branes. The latter are defined in terms of
the geometrical positions of the stacks of D3- and D7-branes on the
$\mathbb{T}^2$, which are the real two-vectors $(a^4,a^5)_i$ for
D3-brane scalars and $(a^4,a^5)_a$ for D7-brane scalars, and
the position vectors are conveniently
complexified to $A = a^4 + U a^5$, for both types of
indices.\footnote{Note that these are the 
proper special-geometry coordinates,
T-dual to the ones used in \cite{Berg:2004ek}.}

In the absence of open string scalars (i.e.\ setting 
$A=0$) 
the three scalars in the three bulk vector multiplets are 
defined as\footnote{The 
conventions of \cite{Berg:2004ek} differ from those of e.g.\ 
\cite{Antoniadis:1996vw}, as explained in \cite{Berg:2004ek}.}
\beqn 
S|_{A=0} = \frac{1}{2 \pi \sqrt{2}} \left(C_{(0)} + i e^{-\Phi_{10}} \right) \ , \quad 
S\, '|_{A=0}  = \frac{1}{2 \pi \sqrt{2}} \left(C_{(4)} + i e^{-\Phi_{10}} {\rm vol(K3)} \alpha'^{-2} \right)\ , 
\eeqn 
where the $C_{(p)}$ are the scalars arising from the RR $p$-forms, 
and $U|_{A=0}  = (G_{45} + i \sqrt{G})/G_{44}$ for the complex 
structure of the $\mathbb{T}^2$. 
It was shown in \cite{Antoniadis:1996vw} that dimensional reduction of 
type I supergravity in the absence of 
open string moduli is described by the prepotential 
$\cf|_{A=0}=SS\, 'U|_{A=0}$. In this limit, 
the gauge coupling of the D7-branes is given by the real part of 
\beqn \label{f7}
f_{\rm 7}|_{A=0} = -i S\, '|_{A=0}  \ , 
\eeqn 
which follows from the Born-Infeld action. 
In the presence of the open string vector multiplets, the fields $S,S\, '$ have to be modified via 
\beqn \label{corrcoord}
S = S|_{A=0} + \frac{1}{8 \pi} \sum_a (a^5)_a A_a \ , \quad 
S\, ' = S\, '|_{A=0} + \frac{1}{8 \pi} \sum_i (a^5)_i A_i\ , \quad  U = U|_{A=0}  \ .
\eeqn
Dimensional reduction (including certain counterterms) 
allows one to deduce the corrected 
K\"ahler potential \cite{Antoniadis:1996vw} 
\beqn \label{kaepot}
\ck &=& - \ln \Big[ (S-\bar S)(S\, '-\bar S\, ')(U-\bar U) \non && 
        \hspace{1cm} - \frac{1}{8 \pi} (S-\bar S) \sum_i (A_i -\bar A_i)^2  
        - \frac{1}{8 \pi} (S\, '-\bar S\, ') \sum_a (A_a -\bar A_a)^2 \Big] \ , 
\eeqn
which follows from a prepotential
\beqn \label{prepotfull}
\cf = SS\, 'U - \frac{1}{8\pi} S \sum_i A_i^2 - \frac{1}{8\pi} S\, ' 
\sum_a A_a^2 \ . 
\eeqn 
The gauge coupling of the D7-brane gauge group is part of the period matrix 
of the $\cn=~2$ Lagrangian defined through this prepotential. To
introduce some more compact
notation, let us collect all fields in projective coordinates 
\beqn 
\left( X^\Lambda/ X^0 \right)
= \left( 1, S, S\, ', U, A_i, A_a \right)  \ . 
\eeqn 
with indices
\beqn 
\Lambda \in \{0,\, ...\, ,3+n_7+n_3\} \; , \quad
i\in\{1,\, ...\, ,n_3\} \; , \quad
a\in \{1,\, ...\, ,n_7\} \; .
\eeqn 
For simplicity, 
we now assume that there is only a single stack of D3- or 
D7-branes: $n_3=1$ and $n_7=1$, hence $\Lambda\in\{0,\, ...\, ,5\}$. 
Then the D7-brane coupling, as derived from supergravity, 
is the imaginary part of the $55$ entry of the period matrix, denoted 
$\cn_{55}$. 
Since we are interested in nonperturbative effects 
on these D7-branes, we focus on
the situation that the gauge group is non-abelian.
The non-abelian coupling reads 
\beqn \label{f7full}
\cn_{55} = S\, '  \ .  
\eeqn 
We therefore recover the $\cn=2$ version of the rho problem, 
since the correction 
present in (\ref{f7full}) through the corrected definition of $S\, '$ 
in (\ref{corrcoord}) does 
not follow from the Born-Infeld (plus Chern-Simons) action, which 
only produces $S\, '|_{A=0}$ as in (\ref{f7}) above
(cf.\ \cite{Hsu:2003cy,Jockers:2004yj}). 
It follows that dimensional reduction of the tree-level 
effective action does not produce the correct supersymmetric Lagrangian. 
We will see that the one-loop 
annulus correction precisely remedies this discrepancy. But first, 
we have to discuss another subtlety, in order to make contact 
with the D3/D7-inflation 
model of \cite{Dasgupta:2002ew,Hsu:2003cy,Hsu:2004hi}, and 
the $\cn=2$ special geometry Lagrangian of 
\cite{Angelantonj:2003zx,D'Auria:2004qv,Antoniadis:1996vw}. 

One of the main features of the Lagrangian 
considered in 
\cite{Dasgupta:2002ew,Hsu:2003cy,Hsu:2004hi,Angelantonj:2003zx,D'Auria:2004qv},
meant to describe the 
string compactification on K3$\times\mathbb{T}^2$ with fluxes, 
is the symplectic transformation 
of the coordinate fields $X^\Lambda$ into a frame where 
no prepotential exists. 
In this symplectic section, the Lagrangian was 
shown to possess a shift symmetry with 
respect to shifting the real parts $A_i+\bar A_i$ of the 
D3-brane coordinates. Therefore, these fields appear as 
exactly flat directions of the 
classical potential, and are candidates for inflaton fields,
provided ``exactly flat'' becomes ``nearly flat'' through some source of symmetry breaking.
Now, our corrections to the D7-brane 
gauge coupling appear to be computed in the frame
where the prepotential (\ref{prepotfull}) exists,
so 
we should apply the symplectic transformation of 
\cite{Angelantonj:2003zx} before any comparison could be made.
%
However, it turns out that this transformation
leaves the gauge coupling 
of the (non-abelian) D7-brane gauge group unchanged, as we now argue.

Let us consider the simplest setting and take the D7-brane 
gauge group to be unbroken, i.e.\ all 
D7-branes in a single stack, and a potentially anomalous 
overall $U(1)$ decoupled. Then, the 
period matrix following from the 
prepotential (\ref{prepotfull}) does not exhibit any terms 
coupling the D7-brane gauge fields to any of the other 
gauge fields. In other words, 
we have $\cn_{5 \Lambda}=0$ for $\Lambda\neq 5$. 
We now apply the symplectic transformation 
\cite{Ceresole:1995jg}
\be
\cn ~\longrightarrow~ \cn\, ' = (B+A\cn) (A-B\cn)^{-1}\ ,
\ee
with the matrices $A$ and $B$ given in \cite{Angelantonj:2003zx}. 
The 7-brane gauge coupling after the transformation is the imaginary part
of 
\be
\cn_{5 5}\!\!\! ' \; = \;
\sum_\Lambda (B+A\cn)_{5\Lambda} (A-B\cn)^{-1}_{\Lambda 5}\ . 
\ee
Due to the simple structure of the matrices $A$ and $B$ we have 
\be
(B+A\cn)_{5\Lambda} = \cn_{5\Lambda} = \cn_{55} \delta_{5\Lambda}\ , 
\ee
where in the last equality we have used the fact that the 7-brane gauge group is non-abelian. 
Thus we arrive at
\be
\cn_{5 5}\!\!\! ' \; 
= \cn_{55} (A-B\cn)^{-1}_{55}
= \cn_{55} \frac{1}{\det(A-B\cn)} \det|A-B\cn|_{55}\ , 
\ee
where  $|A-B\cn|_{55}$ denotes the minor $M_{55}$ of the matrix $A-B\cn$.
Because of $(A-B\cn)_{5\Lambda}=\delta_{5\Lambda}$, 
we can use the expansion formula for determinants
to see that $\det(A-B\cn)=\det|A-B\cn|_{55}$ and 
thus the announced result follows:\footnote{Note 
that for this derivation one only needs that
\be
\cn_{5 \Lambda}=\cn_{55} \delta_{5\Lambda}\ , \ A_{5\Lambda}=\delta_{5\Lambda}\ , \ B_{5\Lambda}=0\ ,
\ee
so that it does not depend on the details of the matrices $A,B$ 
used in \cite{Angelantonj:2003zx} but holds for every symplectic 
transformation not mixing the non-abelian 7-brane gauge fields 
with any of the abelian vectors.
}
\be
\cn_{5 5}\!\!\! ' \; = \cn_{55}\ .
\ee
This shows that the non-abelian D7-brane gauge coupling 
is unaffected by the symplectic transformation, which 
justifies simply adding our correction 
to the entry $\cn_{55}$ of the period matrix in the Lagrangian of 
\cite{Hsu:2003cy,Hsu:2004hi,D'Auria:2004qv}. 
We can now put things together to solve the rho problem, and show 
how the one-loop corrections 
to the gauge coupling constants break the shift symmetry of the $\cn=2$ 
Lagrangian of 
\cite{Hsu:2003cy,Hsu:2004hi,D'Auria:2004qv,Firouzjahi:2003zy}.\footnote{We 
do not want to imply that this is a problem for these papers. 
To construct a viable model for inflation, one 
always has to break the shift symmetry 
at some point. Our one-loop corrections present one way to do so.
An alternative way has been put forward in \cite{renatastalk}, 
which makes use of the misalignment of the D3/D7-branes along the 
lines of \cite{Herdeiro:2001zb}. We thank R.\ Kallosh for discussions 
on this point.} 
As outlined above, the dependence of the gauge couplings on the 
open string scalars arises 
from annulus and M\"obius strip string diagrams, i.e.\ one-loop 
threshold corrections \cite{Bachas:1996zt}. 
The most general expression for the D7-brane gauge coupling constant
can be read off from formula (44) in \cite{Berg:2004ek}, 
valid for all $A_i,\, A_a$ distinct and non-vanishing. 
For the present purpose, it will be sufficient to consider the 
situation where all D7-branes are located 
at the origin, i.e.\ $A_a=0$,\footnote{According to the results of 
\cite{D'Auria:2004qv}, this is the value at a 
minimum of the 3-form flux-induced potential at which $\cn=0$ or $\cn=1$
is preserved.}
which leads to 
\beqn \label{f7b}
f_7 = -iS\, ' - \frac{1}{8 \pi^2} \sum_i \ln \tht_1(A_i,U) + \frac{1}{\pi^2} \ln \eta(U)\ .
\eeqn
Note that the one-loop corrections contain a term that completes 
$S\, '$ to the modified form (\ref{corrcoord}), thus  
solving the rho problem in the ${\cal N}=2$ case. 
Typically, strong coupling effects in the D7-brane gauge group
(for instance instantons) then lead to a dependence of the 
potential on $e^{-{\rm Re}\, f_7}$.
In this way, the additional holomorphic terms in (\ref{f7b})
induce explicit dependence on the real parts $A_i+\bar A_i$ of the 
D3-brane scalars in the potential, and thus lift the shift symmetry of 
the effective Lagrangian.


\section{On warped brane-inflation}

\begin{figure}[h]
\begin{center}
  \resizebox{7cm}{!}{\psfig{figure=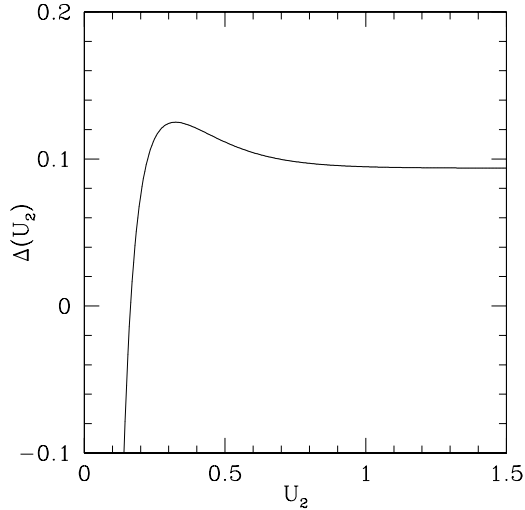,width=6cm}}
\caption{The function $\Delta$ of eq.\ (\ref{r}); 
for positive values the inflaton mass
is lowered by the one-loop corrections to the superpotential.} 
\label{reissdorfkoelsch}
\end{center}
\end{figure}

We now make contact with the KKLMMT model of warped 
brane inflation with only $\cn=1$ 
supersymmetry. This is done by embedding the 
previous $\cn=2$ system of D3- and D7-branes as 
a subsector of an orientifold with $\cn=1$ supersymmetry.
This program can be performed within any $\cn=1$ toroidal 
orientifold based on $\mathbb{T}^6/\mathbb{Z}_{2N}$ or 
$\mathbb{T}^6/(\mathbb{Z}_{2N}\times \mathbb{Z}_{2M})$.\footnote{Note,
however, that in some cases the quadratic contribution in the 
inflaton field drops out of the gauge kinetic function due to 
the extra symmetry at the orbifold point. This does not happen in the 
$\mathbb{T}^6/(\mathbb{Z}_{2}\times \mathbb{Z}_{2})$ model which is  
our main focus in the following.}
All of these models contain 32 D3-branes transverse 
to the compact space, plus 
at least one set of D7-branes wrapped on a 4-cycle, 
defined through an element of the 
orientifold group that acts as the reflection of four circles. 
In that case, the corrections to the gauge kinetic functions as 
discussed above for 
K3$\times \mathbb{T}^2$ can be literally copied to the $\cn=1$ model, 
only subject to more complicated projections on the Chan-Paton 
labels of the charged fields. 
Geometrically, this means that one has to impose the global 
symmetries of the orbifold 
space on the motion of the D3-branes, which was formerly 
unconstrained on $\mathbb{T}^2$. 
All other corrections to the gauge kinetic function of the 
D7-branes, if present, will be independent of the D3-brane 
scalars, and therefore of no concern here.\footnote{Technically, the
reason for this is that only the $\cn=2$ sectors of the 
orientifold depend on the moduli, and it is the special 
$\cn=2$ sector corresponding to the previously mentioned reflection 
along the four directions of the wrapped D7-branes that depends 
on the distance between the D3- and these D7-branes, i.e.\ the inflaton.
The amplitudes arising in this sector are formally identical (up to 
some normalization factors) to the $\cn=2$ case discussed above,
see App.\ B of \cite{Berg:2004ek}.} For 
the simple choice $\mathbb{T}^6/(\mathbb{Z}_{2}\times \mathbb{Z}_{2})$, 
the total modification 
compared to the results for K3$\times \mathbb{T}^2$ boils down to 
nothing but unimportant normalization factors, and a different gauge group. 
To adopt the notation of KKLMMT \cite{Kachru:2003sx}, 
we again locate all the D7-branes at the origin, and 
further concentrate on a single mobile D3-brane, denoting its 
position scalar
by $\phi$. This is the inflaton candidate.\footnote{So 
this $\phi$ is one among the $A_i$ of the previous section.}  
We also rename $S\, '$ to $\rho$ as in the introduction. 
Explicitly, the gauge kinetic function of the D7-branes is then given by 
\beqn 
f_7 = -i\rho - \frac{1}{4 \pi^2} \ln \tht_1(\phi,U) + \, \cdots  \ , 
\eeqn
where the ellipsis indicates terms independent of $\phi$. 
The modulus $U$ is just one of the complex structure moduli 
of the background orientifold space, and assumed to be fixed 
through 3-form fluxes, i.e.\ $U$ 
should be thought of as a function of the quantized flux parameters. 
When gaugino condensation takes place, 
one adds the superpotential 
\beqn 
W_{\rm nonpert} ~\sim~ C e^{-a f_{\rm 7}} ~=~ w(\phi,U) e^{ia\rho} \ . 
\eeqn 
One can now extract the terms that potentially contribute to the inflaton mass 
by expanding $w(\phi,U)$ around some background value for $\phi$. 
For $\phi=1/2$, 
the function is even in $\phi$, and one has 
\beqn 
w(\phi+1/2,U) ~\sim~ \vartheta_1(\phi+1/2,U) ~=~ w_0(U) \left(1 +
a w_2(U) \phi^2 +\, \cdots \right)\ , 
\eeqn
where a factor $a$ has been made explicit for later convenience. 
For general background values, there would also be a linear 
term.\footnote{The value $\phi=0$ 
would not be a consistent choice in the present discussion, 
since then the D3-branes 
and D7-branes sit on top of each other, 
and  new massless states appear. See also \cite{Ganor:1996pe}.}
In the language of KKLMMT  (appendix F of \cite{Kachru:2003sx}) 
this is a concrete identification of their  
parameter $\delta$ as a function of $U$, and thus 
as a function of 
the 3-form flux parameters, as argued above. 
One can now go back to the full scalar potential, insert the 
above superpotential, and find a correction 
to the slow-roll parameter $\eta$ in the form 
\cite{Kachru:2003sx}
\beqn \label{r}
\eta &=& \frac23 \left( 1- \frac{|\cv_{\rm AdS}|}{\cv_{\rm dS}} \Delta(U) \right) \ , 
\eeqn 
where $\cv_{\rm AdS}$ is the value of the potential at the minimum 
before adding the tension of the anti-D3-branes,
$\cv_{\rm dS}$ is the value afterwards, and 
\be
\Delta(U) = -w_2(U) - 2 (w_2(U))^2\ .
\ee
Importantly, $\Delta(U)$ only depends on the quadratic 
coefficient $w_2(U)$ of $w(\phi,U)$, 
and not on the normalization $w_0(U)$, which is much harder 
to obtain. 
From (\ref{r}) one easily infers that $\Delta(U)$ has to be positive 
in order to reach a lower value for $\eta$. 
It turns out that $\Delta(U)>0$ for a 
wide range of values for $U$, cf.\ fig.\ \ref{reissdorfkoelsch}. \\ 

Thus, if the complex structure is 
stabilized in this range, $\eta$ is lowered. 
Determining the actual value of $\eta$ requires
additional input, however; clearly our calculation does not 
capture details of the KKLMMT background, as our derivation of 
the one-loop correction to the gauge kinetic function does not
take into account any effects of the warping or the background 
fluxes. Both can be expected to alter the result at least 
quantitatively, but qualitatively, 
we believe that our correction terms  
survive in the full background. Also, a conclusive 
answer to the question of how quantum corrections 
might help to solve the inflaton 
mass problem cannot be reached without also calculating the 
one-loop corrections to the K\"ahler potential. We will return to 
this problem in the near future \cite{gg}. 
Nevertheless, we consider it a merit of our result that it shows 
how terms quadratic in 
the inflaton field with moduli-dependent coefficients appear in the 
superpotential in an explicit string theory model. This allows 
for moderate fine tuning of $\eta$ 
by choosing flux parameters, as anticipated 
in \cite{Kachru:2003sx}.


\begin{center}
{\bf Acknowledgments} 
\end{center}

We thank Renata Kallosh for 
illuminating insights, helpful discussions and email conversations, and 
Mark Srednicki for discussion. 
It is a pleasure to also thank the organizers of the conferences 
{\it String Phenomenology 2004}, and  
{\it Particles, Strings and Cosmology, PASCOS'04}, where most of the material 
of this article was presented. M.~B.\ was supported by the 
Wenner-Gren Foundations, 
and M.~H.\ by the German Science Foundation (DFG). 
Moreover, the research of M.~B.\ and M.~H.\ was supported in part by 
the National Science Foundation under
Grant No. PHY99-07949. 
The work of B.~K.~was supported by 
the German Science Foundation (DFG) and in part by
funds provided by the U.S. Department of Energy (D.O.E.) 
under cooperative research agreement
$\#$DF-FC02-94ER40818.  


\begin{appendix}

\section{Coupling correction and vertex operators}
\label{appA}

In this appendix we 
use standard vertex operator methods to 
verify our calculation in \cite{Berg:2004ek},
that was performed in the background field formalism. 
The calculation presented here is not fully self-contained, 
and the interested reader should 
consult \cite{Berg:2004ek} and also
\cite{Bain:2000fb} 
for more details on the 
conventions and notation. 
Concretely, we compute a 2-point function 
on an 
annulus with one end on the D7-branes and one on the 
D3-branes (cf.\ fig.\ \ref{figur}). 
This is meant to demonstrate
 that the background field method does indeed give the correct answer 
for the one-loop correction to the D7-brane gauge coupling in the
 presence of dynamical D3-branes. 
We start with\footnote{We are not interested in the overall 
normalization here and do not keep track of it in the following.} 
\be \label{a37}
\langle
\,V_A \, V_A \,\rangle_{\ca_{37}} 
\;\sim\; \int_0^1 \frac{dq}{q} \int_0^q \frac{dz}{z} \int 
\frac{d^4p}{(2 \pi)^4} \sum_{k=0,1} {\rm Tr} \Big[ \theta^k 
V_A(\epsilon_1,p_1;z) 
V_A(\epsilon_2,p_2;1)
q^{L_0} \Big]\ ,
\ee
where the zero-picture open string vertex operators are given by
\be
V_A(\epsilon,p;z) = \lambda\, \epsilon_\mu (\partial X^\mu 
+ i (p \cdot \psi) \psi^\mu)
e^{i p \cdot X(z)}\ . \label{vertex}
\ee
The second vertex operator has been fixed at $z_0=1$ and the 
Chan Paton matrix can be chosen, for instance, as 
\be
\lambda = \frac12 \,{\rm diag}
(\underbrace{1,-1,0,\, ...,0}_{16},\underbrace{-1,1,0,\, ...,0}_{16})\ .
\ee
We use an off-shell prescription here, considering a small
non-vanishing $\delta= p_1 \cdot p_2$.\footnote{Note that this is 
an entirely 
different $\delta$ than the one referred to above in the discussion of 
the KKLMMT model.} 
Without such a prescription, all 2-point functions 
for massless external states
are naively zero due to kinematics. 
The validity of this prescription can be checked by computing a 3-point
function and taking a zero-momentum limit, or by curving the external
space as an IR regulator, cf.\ \cite{Kiritsis:1997hj}. 

Performing a calculation very similar to 
\cite{Bain:2000fb} but including the effect of 
moving the 3-branes away from the 7-branes along the 
torus (cf.\ \cite{gg}) yields
\beqn
\langle
\,V_A \, V_A \,\rangle_{\ca_{37}} 
&\sim& [(p_1 \cdot p_2) (\epsilon_1 \cdot \epsilon_2)
- (p_1 \cdot \epsilon_2) (p_2 \cdot \epsilon_1)] \sum_{k=0,1}
{\rm tr} (\gamma_7^{-k} \lambda_1 \lambda_2) \times 
\non
&& 
\hspace{-1.5cm}  
\int_0^{i \infty} d\tau
  \sum_{\alpha,\beta \atop {\rm even}}  {1\over 2}\, \eta_{\alpha,\beta}\
\frac{1}{4 \pi^4 \tau^2} 
{\thb{\alpha}{\beta}^2 \over \eta^6}
Z^{\alpha,\beta}_{{\rm int},\; k}\,  {\rm tr} 
\left( \Gamma^{(2)}(\vec{a}) \gamma_3^k \right) 
\int_0^{\tau} d\tn\; 
\Big(\langle \psi(z)\psi(1)\rangle \zba{\alpha}{\beta}\Big)^2\ ,
\nonumber
\eeqn
where  $z=e^{2\pi i \tn}$. 
In contrast to \cite{Bain:2000fb} we immediately
omitted the term from the contraction $\langle \partial X 
\partial X \rangle$ because we are only interested in the 
supersymmetric case, where its contribution vanishes after 
summing over spin structures. Finally, we set $\delta=0$ 
everywhere except in the overall factor. This is admissible 
because in the case at hand the integration over $\tn$ does  
not lead to any poles in $\delta$ (cf.\ \cite{Antoniadis:1996vw,gg} 
for more details).
Furthermore, the internal partition function is given by 
\cite{Angelantonj:2002ct,Gimon:1996ay}
\be
Z^{\alpha,\beta}_{{\rm int},\; k} = 
\frac{\thb{\alpha+1/2}{\beta+k/2}(0)\thb{\alpha-1/2}{\beta-k/2}(0)}
{\thb{\;\;\;\;\;0}{1/2+k/2}(0)\thb{\;\;\;\;\;0}{1/2-k/2}(0)}\ ,
\ee
the fermion correlator is \cite{Antoniadis:1996vw}
\be
\langle \psi(z) \psi(1) \rangle \zba{\alpha}{\beta} = i \pi  \,
\frac{\thb{\alpha}{\beta}(\tn) \eta^3}{\vartheta_1(\tn) 
\thb{\alpha}{\beta}(0)} \ , 
\ee
and $\Gamma^{(2)}$ is the sum over winding modes 
\be \label{windings}
\Gamma^{(2)}(\tau,\vec{a}) = \sum_{\vec{n}} 
e^{i \pi \tau (\vec{n} + \vec{a})^{\rm T} G (\vec{n} + \vec{a})}
=\thba{\vec{a}}{\vec{0}}(0,\tau G)\ ,
\ee
with $\thba{\vec{a}}{\vec{0}}$ a genus-two theta function, and  
$\vec{a}$ the position of the 3-brane on the torus.\footnote{More
precisely, the position is $a^4 \vec{e}_1 + a^5 \vec{e}_2$ with 
$\vec{e}_1$ and $\vec{e}_2$ the basis vectors of the torus lattice. Moreover,
in (\ref{windings}) we have set $\alpha'=1/2$.} 
Now, using the identity 
\beqn 
\sum_{\alpha, \beta \atop {\rm even}} 
\eta_{\alpha\beta} \thba{\alpha}{\beta}(\tn)^2
 \thba{\alpha+h}{\beta+g}(0)\thba{\alpha-h}{\beta-g}(0)
= \vartheta_1(\tn)^2 \thba{1/2+h}{1/2+g}(0)\thba{1/2-h}{1/2-g}(0) 
\eeqn 
for $h=g=1/2$
greatly simplifies the integrand as 
all the contributions of string oscillators drop out.
After integration over $\tn$, we arrive at
\beqn \label{a37simp}
&&\hspace{-3cm}
\langle
\,V_A \, V_A \,\rangle_{\ca_{37}} 
\;\sim\; [(p_1 \cdot p_2) (\epsilon_1 \cdot \epsilon_2)
- (p_1 \cdot \epsilon_2) (p_2 \cdot \epsilon_1)] 
\nonumber \\[1mm]
&& \hspace{1cm} \times
\sum_{k=0,1} {\rm tr} (\gamma_7^{-k} \lambda_1 \lambda_2) 
\int_{1/\Lambda^2}^{\infty} \frac{d t}{t} {\rm tr} 
\Big(\thba{\vec{a}}{\vec{0}}(0,i t G) \gamma_3^k\Big)\ .
\eeqn
We changed the variable of integration to
$t=-i\tau$, and 
introduced a UV
cutoff $\Lambda$ in the $t$-integral in (\ref{a37simp}). 
In complete analogy to \cite{Berg:2004ek}, we can perform the traces and 
the integral to obtain
\beqn \label{a37final}
&&\hspace{-3cm}
\langle
\,V_A \, V_A \,\rangle_{\ca_{37}} 
\;\sim\; [(p_1 \cdot p_2) (\epsilon_1 \cdot \epsilon_2)
- (p_1 \cdot \epsilon_2) (p_2 \cdot \epsilon_1)] 
\nonumber \\[1mm]
&&\hspace{1cm}
\times \, \Big( \frac{\Lambda^2}{\sqrt{G}} 
-\ln \left| \frac{\tht_1(A, U)}{\eta(U)} \right|^2 
+ 2\pi U_2 (a^5)^2 \Big) \; ,
\eeqn
where we defined the complexified brane position
\be \label{a}
A = a^4 + U a^5\ .
\ee
The term proportional to $\Lambda^2$ in 
(\ref{a37final}) is independent of the 3-brane scalars and 
drops out of the final result after adding the contribution of 
the 77-annulus and the 7-M\"obius diagrams.
Note, in particular, the difference of (\ref{a}) to the
T-dual variable used in \cite{Berg:2004ek}, which will be
denoted $\tilde A = U a_4-a_5$ in the following. 
Let us compare (\ref{a37final}) to the 
result of \cite{Berg:2004ek} by T-dualizing 
along both directions of the torus (actually along all six internal 
directions, but the K3 directions are not relevant here). 
The background-field result in the 59 picture, from (40) in 
\cite{Berg:2004ek}, reads
\be \label{a95}
\ca_{95}^{{\cal F}^2} \;\sim\; \Big( \Lambda^2 \sqrt{G} 
-\ln \left| \frac{\tht_1(\tilde A, U)}{\eta(U)} \right|^2 
+ 2\pi U_2 a_4^2 \Big) \; .
\ee
If we now apply the rules of T-duality, i.e.\
\be \label{tdual}
U \longrightarrow -\frac{1}{U} \quad , \quad \sqrt{G} \longrightarrow 
\frac{1}{\sqrt{G}} \quad , \quad a_i \longrightarrow a^i\ ,
\ee
which imply
\be
\tilde A \longrightarrow -\frac{1}{U} A\ ,
\ee
we see that (\ref{a95}) is mapped according to 
\beqn \label{trafo}
&&\hspace{-2cm}
\Big( \Lambda^2 \sqrt{G} -\ln \left| \frac{\tht_1(\tilde A, U)}{\eta(U)} 
\right|^2 +2\pi U_2 a_4^2 \Big) \nonumber  \\
&& \longrightarrow\;
 \Big( \frac{\Lambda^2}{\sqrt{G}} 
-\ln \left| \frac{\tht_1(-A/U, -1/U)}{\eta(-1/U)} 
\right|^2 + 2\pi \frac{U_2}{|U|^2} (a^4)^2 \Big) \; .
\eeqn
Using standard 
modular transformations (cf.\ \cite{Kiritsis:1997hj}), 
one easily verifies that (\ref{trafo}) coincides with the bracket in 
(\ref{a37final}). 

As (\ref{a37final}) gives the dependence 
of the one-loop correction to the 7-brane gauge coupling on the 
3-brane scalars, this confirms the results derived in the T-dual picture 
using the background field method in \cite{Berg:2004ek}.

\end{appendix}


\end{document}